\documentclass[12pt]{article}

\usepackage{amsmath,amssymb,amsfonts,amsbsy}
\usepackage{cite}
\usepackage{graphicx}
\usepackage{wrapfig}
\usepackage[font=small,labelfont=bf,labelsep=period]{caption}

\begin{document}
\addcontentsline{toc}{subsection}{{HARD COLLISIONS OF SPINNING PROTONS: HISTORY \& FUTURE}\\
{\it A.D. Krisch}}

\setcounter{section}{0}
\setcounter{subsection}{0}
\setcounter{equation}{0}
\setcounter{figure}{0}
\setcounter{footnote}{0}
\setcounter{table}{0}

\begin{center}
\textbf{HARD COLLISIONS OF SPINNING PROTONS: HISTORY \& FUTURE}

\vspace{5mm}

A.D.~Krisch$^{\,1\,\dag}$

\vspace{5mm}

\begin{small}
  (1) \emph{Spin Physics Center, University of Michigan, Ann Arbor, MI 48109-1040, USA} \\
  $\dag$ \emph{E-mail: krisch@umich.edu Opening Talk: DSPIN-09, JINR, Dubna RU (1-5 Sept 2009)}
\end{small}
\end{center}

\vspace{0.0mm} 

\begin{abstract}
  There will be a review of the history of polarized proton beams,
 and a discussion of the unexpected and still unexplained large 
transverse spin effects found in several high energy proton-proton spin 
experiments at the ZGS, AGS, Fermilab and RHIC. Next, there will be a discussion 
of possible future experiments on the violent elastic 
collisions of polarized protons at IHEP-Protvino's 70 GeV U-70 accelerator 
in Russia and the new high intensity 50~GeV J-PARC at Tokai in Japan.
\end{abstract}

\vspace{7.2mm} 

I will first discuss the violent elastic collisions of unpolarized protons. Fig.~1 shows the 
cross section for proton-proton elastic scattering plotted against a scaled $%
P_{t}^{2}$ variable that was proposed in 1963~\cite{1} and 1967~\cite{2} following Serber's~\cite{3} optical model.
This plot is from updates by Peter Hansen and me~\cite{4,5}. Notice that at
small $P_{t}^{2}$ the cross-section drops off with a slope of about
10~(GeV/c)$^{-2}$. Fourier transforming this slope gives the size and shape
of the proton-proton interaction in the diffraction peak; it is a Gaussian
with a radius of about 1~Fermi. At medium $P_{t}^{2}$ there is a component
with a slope of about 3~(GeV/c)$^{-2}$; however, this component disappears
rapidly with increasing energy. At lab energies of a few TeV it has totally
disappeared; then, there is a sharp destructive interference between the
small-$P_{t}^{2}$ diffraction peak and the large-$P_{t}^{2}$ hard-scattering
component. Since the diffraction peak is mostly \textit{diffractive}, its
amplitude must be mostly imaginary, as has been experimentally verified.
Hence, the sharp destructive interference implies that the large-$P_{t}^{2}$
component is also mostly imaginary; thus, it is probably mostly \textit{%
diffractive}. This large-$P_{t}^{2}$ component is probably the
elastic \textit{diffractive} scattering due to the \textit{direct} interactions of
the proton's constituents; its slope of about 1.5~(GeV/c)$^{-2}$ implies
that these \textit{direct} interactions occur within a Gaussian-shaped
region of radius about 0.3~Fermi.

Since the medium-$P_{t}^{2}$ component disappears at high energy, it is
probably the \textit{direct elastic scattering} of the two protons. This
view is supported by the experimental fact that proton-proton elastic
scattering is the only exclusive process that still can be precisely
measured at TeV energies. To understand this, note that \textit{direct}
elastic scattering and all other exclusive processes must compete with each
other for the total p-p cross-section, which is less than 100~millibarns. At
TeV energies, there are certainly more than 10$^{5}$ exclusive channels in
this competition; thus, each channel has an average cross section of less
than 1~microbarn. Moreover, since the medium-$P_{t}^{2}$ elastic component
does not interfere strongly with either the large-$P_{t}^{2}$ or small-$%
P_{t}^{2}$ components, its amplitude is probably real. Also note that the
large-$P_{t}^{2}$ component intersects the cross section axis at about $10^{-5}$
below the small-$P_{t}^{2}$ \textit{diffractive} component.

\begin{figure}[b!]
  \begin{tabular}{@{}l@{\hspace{5mm}}r}
	\begin{minipage}{87mm}    
	\centering
    \includegraphics[width=87mm,height=157mm]{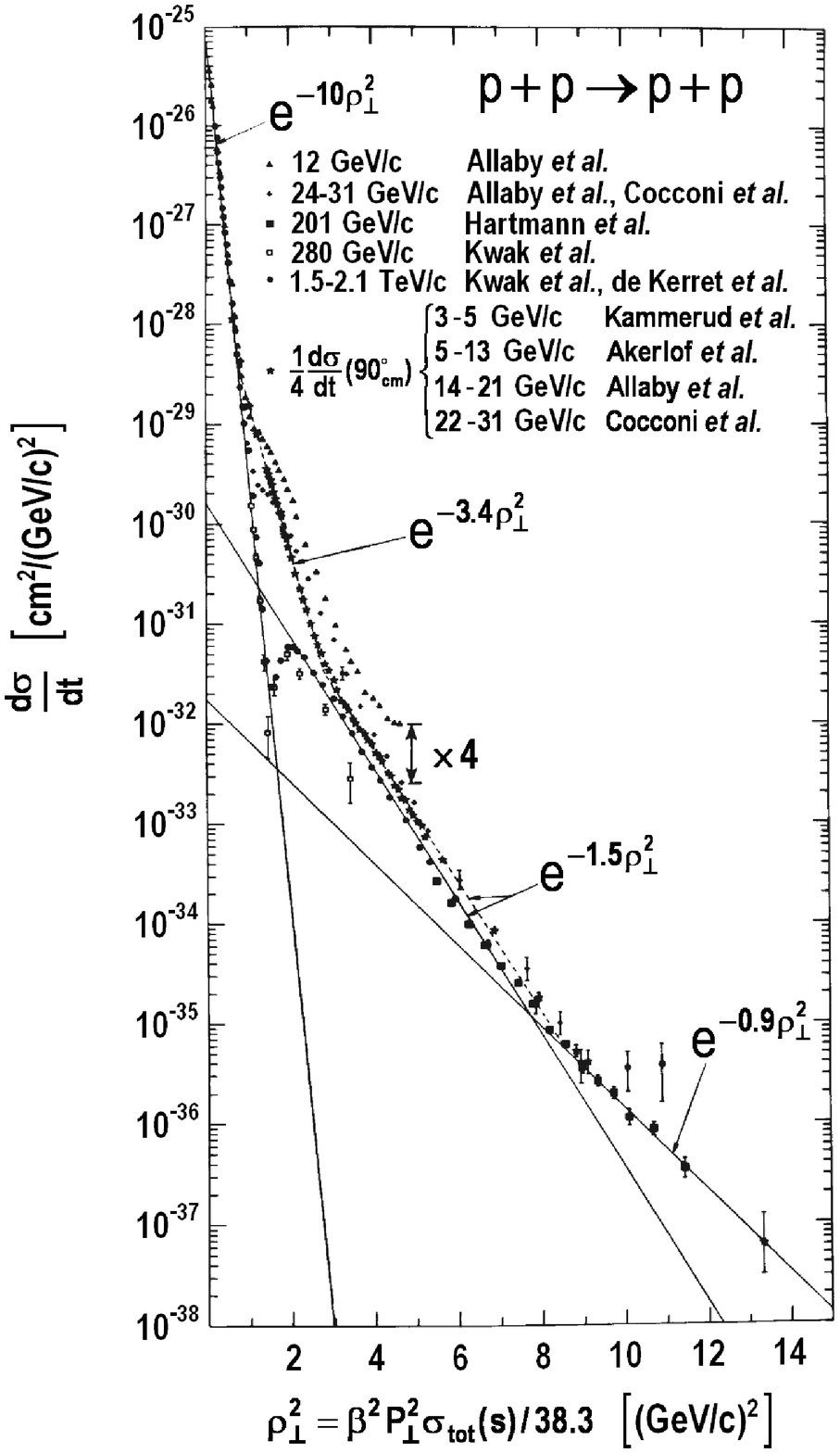} 
    \caption{Proton-proton elastic cross-sections plotted vs. the scaled 
$P_{t}^{2}$ variable~\cite{4,5}. The 12 GeV/c Allaby \textit{et al.} data were not 
corrected for $90_{cm}^{\circ}$ particle identity effects.} 
	\label{fig1}
	\end{minipage}
    &
	\begin{minipage}{68mm} 
	\centering
    \includegraphics[width=67mm,height=80mm]{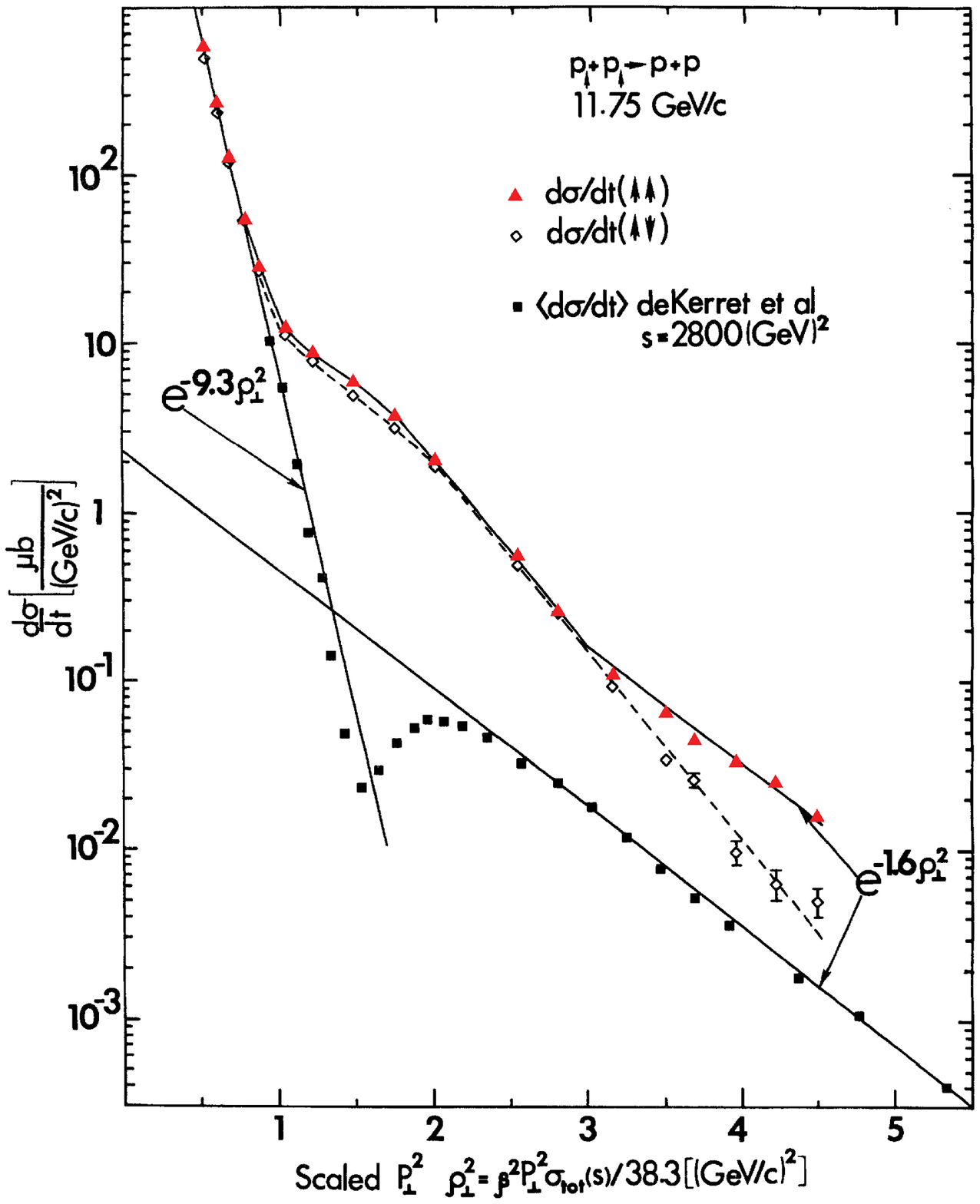} 
    \caption{The proton-proton elastic cross section near 12~GeV in pure 
initial spin states plotted vs. the scaled $P_{t}^{2}$-variable~\cite{11}.}  
	\label{fig2}
	\vspace*{4mm} 
	\centering
    \includegraphics[width=67mm,height=52mm]{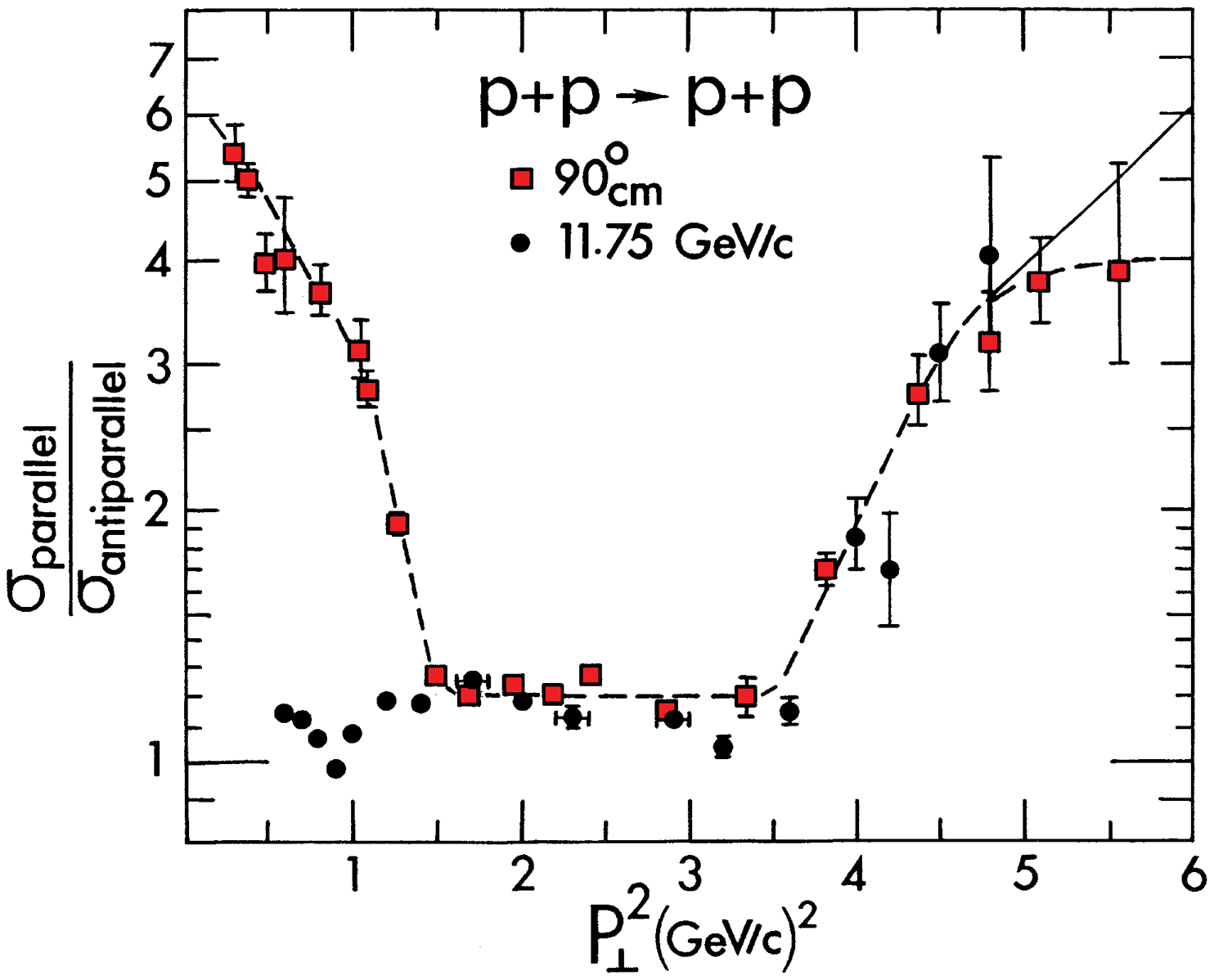}
    \caption{The measured spins-parallel / spins-antiparallel elastic cross-section ratio $(\sigma_{\uparrow \uparrow}/\sigma_{\uparrow\downarrow})$ plotted vs. $P_{t}^{2}$~\cite{12}.}
	\label{fig3}
	\end{minipage}
  \end{tabular}
\end{figure}

An earlier version of Fig.~1 got me started in the spin business. In 1966,
we measured p-p elastic scattering at the ZGS at exactly 
$90_{cm}^{\circ }$ from 5 to 12~GeV~\cite{6}; the sharp slope-change, shown by
the stars, was apparently the first direct evidence for constituents in the
proton. Dividing these~$90_{cm}^{\circ }$ p-p elastic cross sections by 4 
(due to the protons' particle identity) made all then existing proton-proton 
elastic data, above a few GeV, fit on a single curve~\cite{2}. During a 1968
visit to Ann Arbor, Prof. Serber informed me that, by dividing the 
$90_{cm}^{\circ }$ points by 4, I had made an assumption about the ratio of 
the spin singlet and triplet p-p elastic scattering amplitudes. I was
astounded and said that I knew nothing about spin and certainly had not
measured the spin of either proton. He said with a smile that both
statements might be true; nevertheless, my nice fit required this
assumption. Prof.~Serber, as usual, spoke quietly; however, as a student, 
I had learned that he was almost always right. Thus, I looked for data on 
proton-proton elastic scattering data, in the singlet or triplet spin states, 
above a few GeV. I found that none existed and decided to try to polarize the protons in the ZGS.

At the 1969 New York APS Meeting, I learned that EG\&G was the
representative for a new polarized proton ion source made by ANAC in New
Zealand. I discussed this with my long-time colleague, Larry Ratner, and
then with Bruce Cork, Argonne's Associate Director, and Robert Duffield,
Argonne's Director. They apparently decided it was a good idea; Duffield
soon hired me as a consultant to Argonne at \$100 per month. In 1973, after
a lot of hard work by many people, the ZGS accelerated the world's first
high energy polarized proton beam~\cite{7}.

The ZGS needed some hardware to overcome both intrinsic and imperfection
depolarizing resonances. Fortunately, both types of resonances were fairly
weak at the 12 GeV ZGS, which was the highest energy weak focusing accelerator 
ever built. All higher energy accelerators wisely use strong focusing~\cite{8}, which
makes the depolarizing resonances much stronger. If we had first tried to
accelerate polarized protons at a strong focusing accelerator, such as the
AGS, we probably would have failed and abandoned the polarized proton beam
business. Fortunately, it worked at the weak focusing ZGS~\cite{7,9}, and experiments~\cite{10} 
soon showed that the p-p total cross-section had significant
spin dependence; this surprised many people, including me.

Figure~2 shows our perhaps most important result~\cite{11} from the ZGS
polarized proton beam. The 12~GeV proton-proton elastic cross section in
pure initial spin states is plotted against the scaled $P_{t}^{2}$-variable;
in the diffraction peak the spin-parallel and spin-antiparallel
cross-sections are essentially equal to each other and to the unpolarized data 
from the CERN ISR at $s=2800~GeV^{2}$ ; thus, in small-angle 
\textit{diffractive} scattering, the protons in different spin states (and
at different energies) all  have about the same cross-section. 
The medium-$P_{t}^{2}$ component, which still exists near 12~GeV, has only a small spin
dependence; again note that it has totally disappeared at 2800~GeV$^{2}$.
However, the behavior of the large-$P_{t}^{2}$ hard-scattering component was
a great surprise. When the protons' spins are parallel, they seem to have
exactly the same behavior as the much higher energy unpolarized ISR data;
however, when their spins are antiparallel their cross-section drops with
the medium-$P_{t}^{2}$ component's steeper slope. When this data first
appeared in 1977 and 1978, people were totally astounded; most had thought
that spin effects would disappear at high energies. In the years following,
many theoretical papers tried to explain this unexpected behavior; none were
fully successful. In particular, the theory that is now called QCD, has
been unable to deal with this data; Glashow once called this experiment "..the
thorn in the side of QCD". In his summary talk at Blois 2005, Stan Brodsky
called this result "..one of the unsolved mysteries of Hadronic Physics".

I learned something important from questions during two 1978 seminars about this
result. Two distinguished physicists, Prof.~Weisskopf at CERN and then
Prof.~Bethe at Copenhagen a week later, asked the same question apparently
independently. Each said that our big spin effect at large-$P_{t}^{2}$ was
quite interesting; but at 12~GeV, the spins-parallel/spins-antiparallel 
ratio was only large near $90_{cm}^{\circ }$, where particle identity was 
important for p-p scattering. They asked: How could one be sure that our 
large spin effect was due to hard-scattering at large-$P_{t}^{2}$, rather 
than particle identity near $90_{cm}^{\circ }$?  One would be foolish to 
ignore the comments of two such distinguished theorists; moreover, they were related 
to Prof.~Serber's comment 10~years earlier. 

It seemed that their question could not be answered theoretically. 
Thus, we tried to answer it experimentally with a second ZGS experiment, which  
varied $P_{t}^{2}$ by holding the p-p scattering angle fixed at exactly 
$90_{cm}^{\circ }$, while varying the energy of the proton beam. This 
$90_{cm}^{\circ }$ p-p elastic fixed-angle data~\cite{12} is plotted against 
$P_{t}^{2}$ in Fig.~3, along with the fixed-energy data~\cite{11} of Fig.~2. There are
large differences at small $P_{t}^{2}$, where the $90_{cm}^{\circ }$ data
are at very low energy; however, above $P_{t}^{2}$ of about 1.5~(GeV/c)$^{2}$, 
the two sets of data fall right on top of each other. The point at 
$P_{t}^{2}$ = 2.5~(GeV/c)$^{2}$, where the ratio is near 1, is just as much
at $90_{cm}^{\circ }$, as the 5~(GeV/c)$^{2}$ point, where the ratio is 4.
This data apparently convinced Profs.~Bethe and Weisskopf that the
large spin effect was not due to $90_{cm}^{\circ}$ particle identity and it 
was a large-$P_{t}^{2}$ hard-scattering effect.

\begin{wrapfigure}[20]{R}{77.5mm}
  \centering 
  \vspace*{-5mm} 
  \includegraphics[width=73mm]{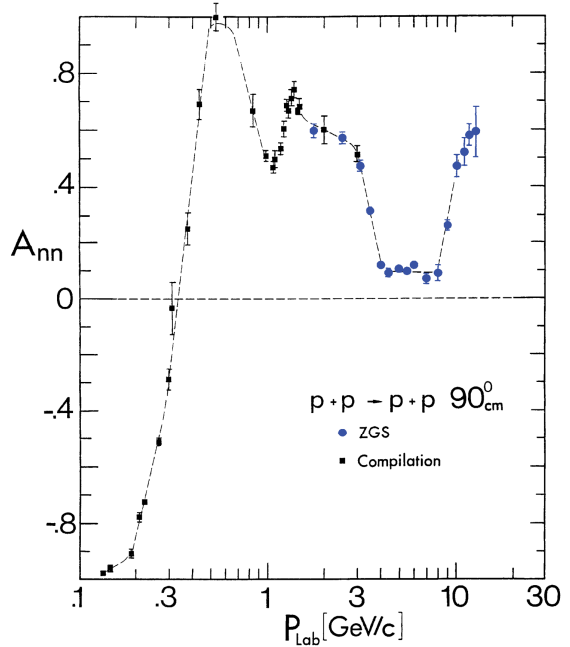}
  \caption{$A_{nn}\equiv(\sigma_{\uparrow \uparrow}-\sigma_{\uparrow\downarrow})/ 
(\sigma_{\uparrow \uparrow}+\sigma_{\uparrow\downarrow})$ is plotted 
against $P_{Lab}$.~\cite{5}}
  \label{fig4}
\end{wrapfigure}

Figure~4 shows $A_{nn}$ for p-p elastic scattering 
plotted against the lab momentum, $P_{Lab}$~\cite{5}; it includes the ZGS data from
Fig.~3 plus some lower energy data obtained from Willy~Haeberli, who is an
expert on low energy p-p spin experiments. At the lowest momentum (near $T=10
$~MeV) $A_{nn}$ is very close to $-1$; thus, two protons with parallel spins can 
never scatter at $90_{cm}^{\circ }$. Next $A_{nn}$ goes rapidly to $+1$; 
then protons with antiparallel spins can never scatter at $90_{cm}^{\circ}$. 
Then at medium energy, there are some oscillations that were
once thought to be due to dibaryon resonances, but may be due to the
onset of $N^{\ast}$ resonance production. In the ZGS region, $A_{nn}$ first
drops rapidly; it is next small and constant over a large range; it then
rises rapidly to 0.6. These huge and sharp oscillations of $A_{nn}$ seem
impressive.

I now turn to funding. In 1972 the AEC had agreed to shut down the 
ZGS in 1975 to get funding for PEP at SLAC. When the unique ZGS polarized
beam started operating in 1973, the wisdom of this decision was questioned;
AEC then set up a committee which extended ZGS operations through 1977.
A second committee in 1976 extended operations of the 
ZGS polarized beam until 1979~\cite{13}. Henry Bohm, the President of AUA, which
operated Argonne, asked ERDA, which had replaced AEC, to set up a third committee to 
again extend ZGS running. However, OMB objected, so there was no third committee; 
nevertheless, his efforts had some benefit. When James Kane, of ERDA, responded 
negatively to Dr. Bohm, his justification was that it might now 
be possible to accelerate polarized protons in a strong focusing accelerator 
such as the AGS; morover, Dr. Kane officially copied me on his letter.

We had started interacting with Ernest Courant and others at Brookhaven
about polarizing the AGS, first at a 1977 Workshop in Ann Arbor~\cite{14}. 
Then at a 1978 Polarized AGS Workshop at Brookhaven~\cite{15},
Brookhaven's Associate Director, Ronald Rau, asked me for a copy 
of Dr. Kane's letter. He used it to convince William Wallenmeyer, the long-time 
Director of High Energy Physics at AEC, ERDA and DoE, to provide about 
\$8~Million to Brookhaven, and about \$2~Million split between Michigan, 
Argonne, Rice and Yale, for the challenging project of accelerating polarized 
protons in the strong-focusing AGS, and later in the 400~GeV ISABELLE collider. 
ISABELLE was canceled in 1983, but was later reborn as RHIC, which is now colliding 
250~GeV polarized protons.

It was far more difficult to accelerate polarized protons in the strong
focusing AGS than in the weak focusing ZGS. The strong focusing principal,
invented by Courant, Livingston and Snyder~\cite{8}, made possible all
modern large circular accelerators by using alternating quadrupole magnetic
fields to strongly focus the beam and thus keep it small. Unfortunately,
these strong quadrupole fields were very good at depolarizing protons. To
accelerate polarized protons to 22~GeV at the AGS, one had to overcome 45
strong depolarizing resonances. This required: some very challenging
hardware; significantly upgrading the AGS controls; and spending lots of
time individually overcoming the 45 depolarizing resonances. Michigan built
12 ferrite quadrupole magnets to allow the AGS to overcome
its 6 intrinsic resonances by rapidly jumping its vertical betatron tune
through each resonance. Brookhaven was building their 12 power
supplies; but each power supply had to provide 1500~Amps at 15,000~Volts (about
22~MW) during each quadrupole's 1.6~$\mu$sec rise-time. Overcoming the many
imperfection depolarizing resonances (occurring every 520~MeV) required
programming the AGS's 96 small correction dipole magnets to form a horizontal
B-field wave of 4 oscillations at the instant when the proton energy passed through 
the $G\gamma =4$ inperfection resonance. Then, about 20~msec later in the AGS cycle, 
when $G\gamma$ was 5, the 96 magnets had a horizontal B-field wave with 5 oscillations,
etc. ($G=1.79285$ is the proton's anomalous magnetic moment, while $\gamma=E/m$.).

After all this hardware was installed, an even larger problem was tuning the
AGS. In 1988, when we accelerated polarized protons to 22~GeV, we needed 7
weeks of exclusive use of the AGS; this was difficult and expensive. Once a
week, Nicholas Samios, Brookhaven's Director, would visit the AGS Control
Room to politely ask how long the tuning would continue and to note 
that it was costing \$1 Million a week. Moreover, it was soon clear that,
except for Larry Ratner (then at Brookhaven) and me, no one could tune
through these 45 resonances; thus, for some weeks, Larry and I worked
12-hour shifts 7-days each week.  After 5 weeks Larry
collapsed. While I was younger than Larry, I thought it unwise to try to
work 24-hour shifts every day. Thus, I asked our Postdoc, Thomas Roser, who
until then had worked mostly on polarized targets and scattering
experiments, if he wanted to learn accelerator physics in a hands-on way for
12~hours every day. Apparently, he learned well, and now leads Brookhaven's
Collider-Accelerator Division.

One benefit from this difficult 7-week period~\cite{16,7} was learning that
our method of individually overcoming each resonance, which had worked so
well at the ZGS~\cite{9,7}, might work at the AGS, but would not be practical at 
higher energy accelerators. This lesson helped to launch our
Siberian snake programs at IUCF~\cite{17,7} and then SSC~\cite{18,19}.

In the 1980's, a new proton collider, the SSC, was being planned; it was to
have two 20~TeV proton rings each about 80~km in circumference. Owen
Chamberlain and Ernest Courant encouraged me to form a collaboration to
insure that polarized protons would be possible in the SSC. We
first organized a 1985 Workshop in Ann Arbor, with Kent Terwilliger. This
Workshop~\cite{18} concluded that it should be possible to accelerate and
maintain the polarization of 20~TeV protons in the SSC, but only if the new
Siberian snake concept of Derbenev and Kondratenko \cite{20} really worked;
otherwise, it would be totally impractical. Recall that it took 49 days to
correct the 45 depolarizing resonances at the AGS, about one per day. Each
20~TeV SSC ring would have about 36,000 depolarizing resonances to correct. 
Moreover, these higher energy resonances would be much stronger and harder to correct; 
but even at one per day, this would require about 100~years of tuning 
for each ring. The Workshop also concluded that one must prove experimentally 
that the \textit{too-good-to-be-true} Siberian snakes really
worked; otherwise, there would be no approval to install the 26 Siberian 
snakes needed in each SSC ring.

Indiana's IUCF was then building a new 500~MeV synchrotron
Cooler Ring~\cite{7}. Some of us Workshop participants then collaborated with Robert
Pollock and others at IUCF to build and test the world's first Siberian
snake in the Cooler Ring. We brought experience with synchrotrons and high
energy polarized beams, while the IUCF people brought experience with low
energy polarized beams and the CE-01 detector, which was our polarimeter. In
1989, we demonstrated that a Siberian snake could easily overcome a strong
imperfection depolarizing resonance~\cite{17,7}. For 13 years we
continued these experiments and learned many things about spin-manipulating
polarized beams. After the IUCF Cooler Ring  shut down in 2002, this polarized proton 
and deuteron beam experiment program was continued at the 3 GeV COSY in Juelich, Germany 
from 2002-2009~\cite{7}.

In 1990 we formed the SPIN Collaboration and submitted to the SSC: Expression of
Interest EOI-001~\cite{19}. It proposed to accelerate and store polarized protons 
at 20~TeV; and to study spin effects in 20~TeV p-p collisions. It was submitted a week 
before the deadline, which made it SSC EOI-001. Thus, we made the 
first presentation to the SSC's PAC before a huge audience that included many 
newspaper reporters and TV cameras. Perhaps partly
due to this publicity, we were soon \textit{partly} approved by SSC
Director Roy Schwitters. By \textit{partly} I mean that he decided to add 26
empty spaces for Siberian snakes in each SSC Ring; each space was about 20~m
long, which added about 0.5~km to each Ring. Unfortunately, the SSC was
canceled around 1993, before it was finished, but after \$2.5~Billion was
spent. Nevertheless, our detailed studies of the behavior and
spin-manipulation of polarized protons at IUCF and COSY helped in developing
polarized beams around the world: Brookhaven now has 250~GeV polarized
protons in each RHIC ring~\cite{21,7}; perhaps someday CERN's 7~TeV LHC might
have polarized protons.

We eventually accelerated polarized protons to 22~GeV in the AGS~\cite{16,7} 
and obtained some $A_{nn}$ data~\cite{22,23,7} well above the ZGS energy of
12~GeV; but we never had enough polarized-beam \textit{data-time} to get precise $A_{nn}$ data at 
high-$P_{t}^{2}$. But, during tune-up runs for the $A_{nn}$ experiment, 
we used the unpolarized AGS proton beam to test our polarized proton target 
and double-arm magnetic spectrometer by\ measuring $A_{n}$ in 28~GeV
proton-proton elastic scattering; this data resulted in an interesting surprise.
Despite QCD's inability to explain the big $A_{nn}$ from the ZGS, our QCD
friends had made a firm prediction that the one-spin $A_{n}$ must go to 0;
they also said this prediction would become more firm at higher energies and in more
violent collisions. But above $P_{t}^{2}$ = 3~(GeV/c)$^{2}$, $A_{n}$ instead
began to deviate from 0 and was quite large at $P_{t}^{2}$ = 6~(GeV/c)$^{2}$. 
This led to more controversy~\cite{23}; some QCD supporters said that
our $A_{n}$ data must be wrong.

Experimenters take such accusations seriously. Thus, we started preparing an
experiment that could study $A_{n}$ at high-$P_{t}^{2}$ with better
precision. Our spectrometer worked well, but we could only use about 0.1\%
of the AGS beam intensity, because a higher intensity beam would heat our
Polarized Proton Target (PPT) and depolarize it. Thus, we started building a
new PPT~\cite{24,7} that could operate with 20 times more beam intensity; this required 
$^{4}$He evaporation cooling at 1~K, which has much more cooling power than 
our earlier $^{3}$He evaporation PPT at 0.5~K. However, to maintain a target polarization
near 50\% at 1~K required increasing the B-field from 2.5 to
5~Tesla. Thus, we ordered a 5~T superconducting magnet from
Oxford Instruments, with a B-field uniformity of a few $10^{-5}$ over the
PPT's 3~cm diameter volume. We also obtained a Varian 20~W at 140~GHz
Extended Interaction Oscillator; it was apparently the highest power 140~GHz
microwave source available. As the PPT assembly started in 1989, we
worried that, if the PPT model was wrong, the polarization might be only 10\%; 
but we were very lucky; it was 96\%~\cite{24,7}.

\begin{figure}[b!]
  \begin{tabular}{@{}l@{\hspace{5mm}}r}
	\begin{minipage}{77.5mm}    
	\centering
    \includegraphics[width=72mm]{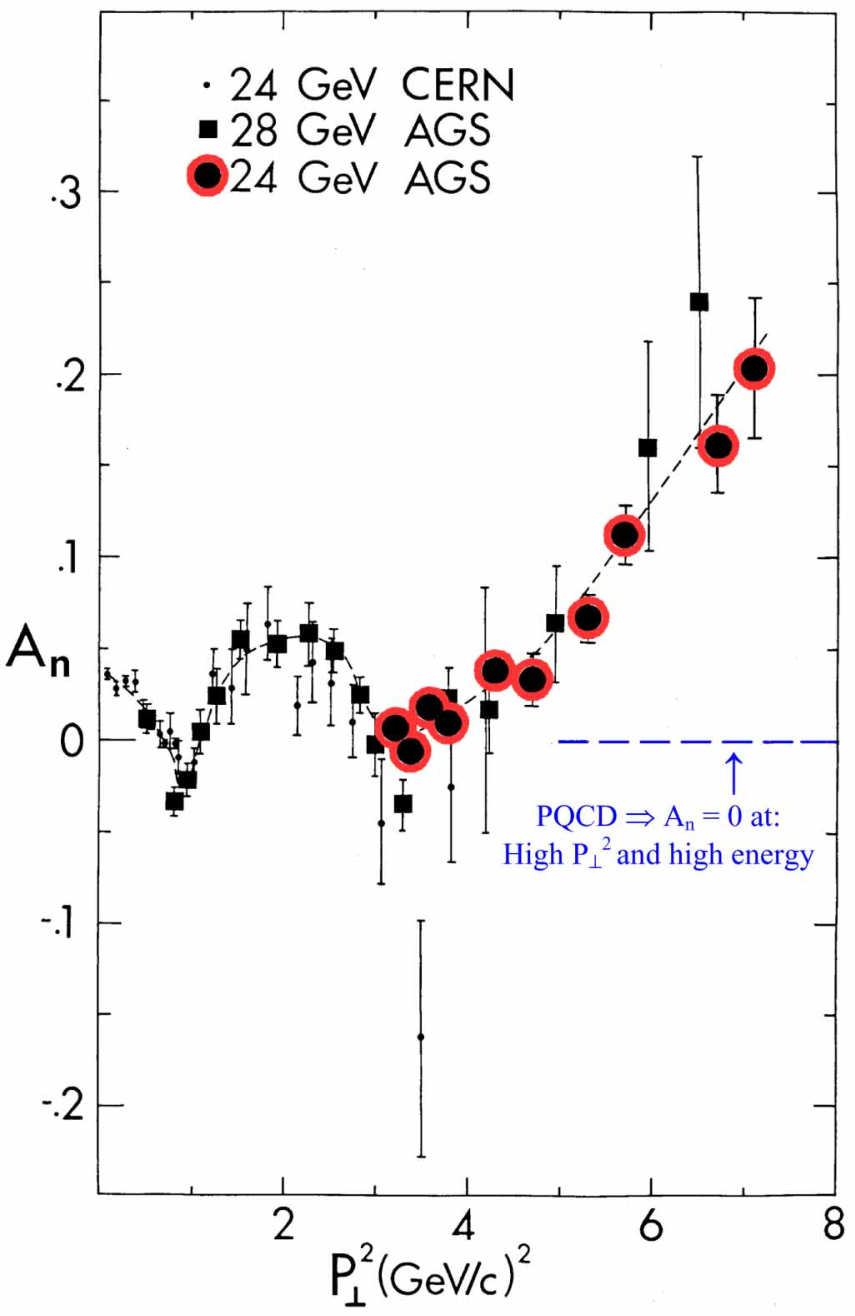} 
	\end{minipage}
    &
	\begin{minipage}{77.5mm} 
	\centering
    \caption{(Left) $A_{n}\equiv(\sigma_{\uparrow}-\sigma_{\downarrow})/ 
(\sigma_{\uparrow}+\sigma_{\downarrow})$ is plotted vs. $P_{t}^{2}$ for p-p elastic scattering~\cite{25}.} 
	\label{fig5}
    \caption{(Bottom) The inclusive $\Lambda$ polarization is plotted against $P_{t}$~\cite{26}.}  
	\label{fig6}
	\vspace*{5mm}
    \includegraphics[width=75mm]{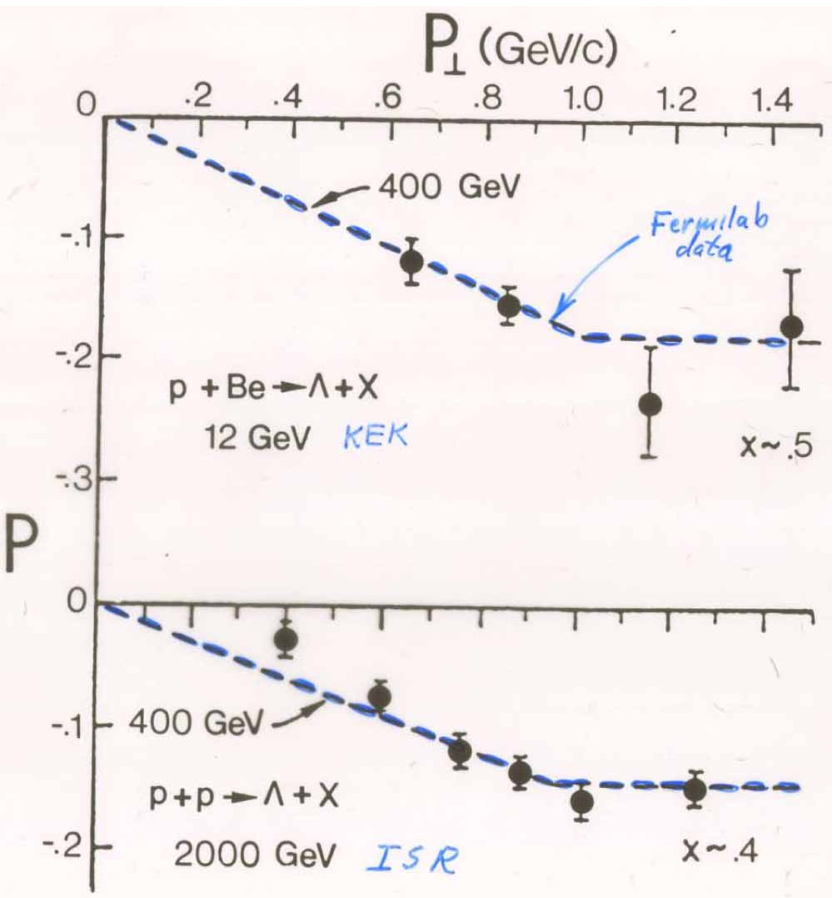} 
	\end{minipage}
  \end{tabular}
\end{figure}

Moreover, the target polarization averaged 85\% for a 3-month-long run with
high-intensity AGS beam in early 1990. As shown in Fig.~5, this let us
precisely measure $A_{n}$ at even larger $P_{t}^{2}$. When these precise new
data were published~\cite{25}, some theorists seemed quite unhappy; they
still believed the QCD prediction that $A_{n}$ must go to 0, but they now
refused to state at what $P_{t}^{2}$ or energy this prediction would become
valid. They also now said that QCD might not work for elastic scattering,
which they now considered less fundamental than inelastic scattering, where
they said QCD should work. Thus, one result of our experiments was to make
both elastic scattering experiments and spin experiments unpopular in some
circles.

Experimenters had also started doing inclusive polarization experiments 
at Fermilab. Figure~6 shows the 400~GeV inclusive hyperon
polarization experiments from the 1970s and 1980s, led by Pondrum, Devlin, Heller and
Bunce~\cite{26}; it clearly shows a small polarization at small $P_{t}$
and a larger polarization at larger $P_{t}$ . Moreover, their data is
consistent with 12~GeV data from the KEK PS and with 2000~GeV data from the
CERN ISR. These data do not support the QCD prediction that
inelastic spin effects disappear at high energy or high $P_{t}^{2}$.

\begin{wrapfigure}[25]{R}{77.5mm}
  \centering 
  \vspace*{-3mm} 
  \includegraphics[width=73mm]{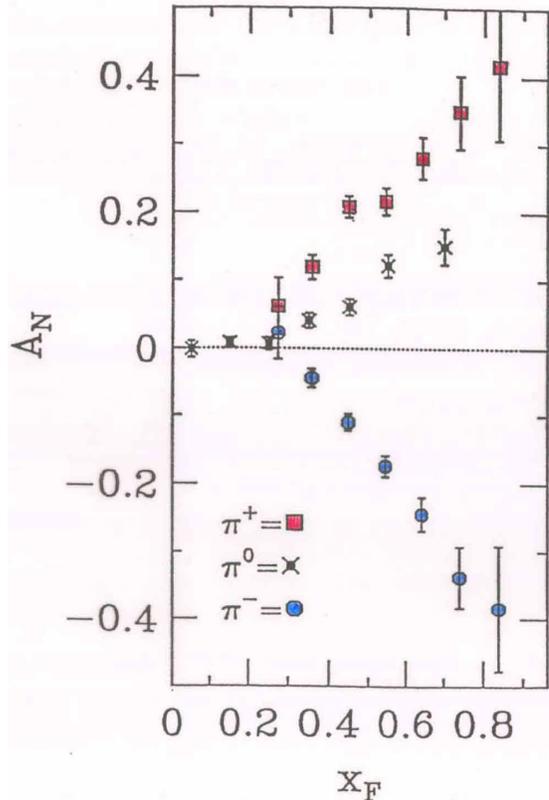}
  \caption{$A_{n}$ for inclusive $\pi$-meson production is plotted vs. $X_F$~\cite{27}.}
  \label{fig7}
\end{wrapfigure}

Another group at Fermilab, led by Yokosawa, developed a \textit{secondary} polarized proton 
beam using the polarized protons from polarized hyperon decay. The beam's
intensity was only about $10^{5}$ per second, but its polarization was about 50\%
and its energy was about 200~GeV. They obtained some nice $A_{n}$ data on
inclusive $\pi$ meson  production~\cite{27}, which are shown in
Fig.~7. The $A_{n}$ values for $\pi^{+}$ and $\pi^{-}$ mesons
are both large but with opposite signs, while $A_{n}$ for the $\pi^{0}$ data is
50\%~smaller and is positive. These 200 GeV data provide little support for QCD.

We tried to measure spin effects in very high energy p-p scattering at UNK,
which IHEP-Protvino started building around 1986~\cite{7}; IHEP and Michigan signed
the NEPTUN-A Agreement in 1989. Michigan's main contribution was a 12 Tesla
at 0.16 K Ultra-cold Spin-polarized Jet~\cite{7}. UNK's circumference would be 21 km
with 3 rings: a 400 GeV warm ring and two 3 TeV superconducting rings; its 
injector was IHEP's existing 70~GeV accelerator, U-70~\cite{7}. By 1998 the UNK
tunnel and about 80\% of its 2200 warm magnets were finished, and 70~GeV
protons were transferred into its tunnel with 99\% efficiency. However,
progress became slower each year due to financial problems; in 1998
Russia's MINATOM placed UNK on long-term standby~\cite{7}. 

IHEP Director, A.A. Logunov, had earlier suggested moving our experiment to IHEP's existing 70 GeV
U-70 accelerator. By March 2002 the resulting SPIN@U-70 Experiment on 70~GeV 
p-p elastic scattering at high $P_{t}^{2}$ was fully installed~\cite{7}, except for our detectors
and Polarized Proton Target (PPT)~\cite{24,7}. However, just before our 4 tons of
detectors, electronics and computers were to be shipped, the US Government
suspended the US-Russian Peaceful Use of Atomic Energy Agreement started by
President Eisenhower in 1953. Nevertheless, DoE asked us to send the 
shipment, since under the terms of the PUoAE Agreement, the experiment should 
have been done exactly as planned; DoE faxed us a copy of the Agreement. Thus, we 
sent the shipment. It arrived at Moscow airport on March 11, 2002, where it was impounded 
for 8 months before it was returned to Michigan.
 
Despite this problem, we remain friends with our IHEP colleagues and there have been four SPIN@U-70 
test runs using Russian detectors and an unpolarized target; we
participated in the November 2001 and April 2002 runs. We hope that the 
US-Russian PUoAE Agreement will soon be restarted so that the SPIN@U-70 experiment 
can continue and measure $A_{n}$ at $P_{t}^{2}$ near 12~(GeV/c)$^{-2}$. However, if this 
international problem continues, we may try to do a similar experiment at Japan's new 
high-intensity 50 GeV proton accelerator, J-PARC~\cite{7} that is now starting to run. 
If J-PARC could accelerate polarized protons to 50 GeV, then one could study the large and 
mysterious elastic spin effects in both $A_{n}$ and $A_{nn}$ for the first time in decades. 

\begin{figure}[ht!]
  \centering
  \includegraphics[width=\columnwidth]{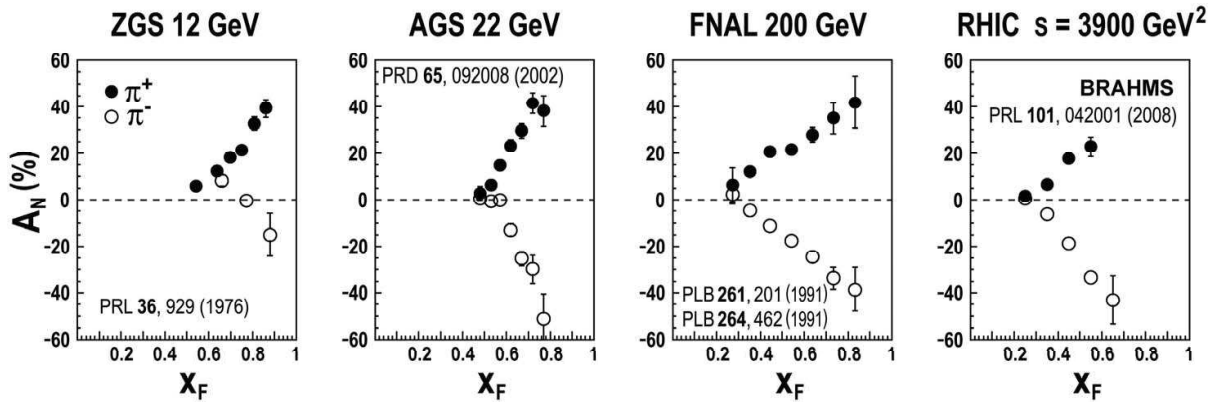} 
  \caption{Inclusive pion asymmetry in proton-proton collisions~\cite{27}.} 
  \label{fig8}
\end{figure}

To summarize, for the past 30 years QCD-based calculations have continued to disagree with the 
ZGS 2-spin and AGS 1-spin elastic data, and the ZGS, AGS, Fermilab and now RHIC~\cite{28} inclusive data. 
To be specific:

  * These large spin effects do not go to zero at high-energy or high-$P_{t}$, as was predicted.

  * No QCD-based model can  yet explain simultaneously all these large spin effects.\linebreak 
There is a BASIC PRINCIPLE OF SCIENCE: 

  * If a theory disagrees with reproducible experimental data, then it must be modified.\linebreak  
Precise spin experiments could provide experimental guidance for the required modification of the theory of Strong Interactions. New experiments at higher energy and higher $P_{t}$ on the proton-proton elastic cross-section's: $d\sigma/dt$, $A_{nn}$ and $A_{n}$ could provide further guidance for these modifications, just as the RHIC inclusive $A_{n}$ experiment~\cite{28} confirmed the earlier Fermilab experiments~\cite{27}.  Elastic scattering is especially important because:

  * It is the only exclusive process large enough to be measured at TeV energy.\linebreak
This is probably because proton-proton elastic scattering is dominated by the diffraction due to the millions of inelastic channels that compete for the total cross-section of only about 100 milibarns at TeV energies. Many people may have forgotten this simple but essential geometrical approach~\cite{1}, which I learned from Prof. Serber's optical model in 1963~\cite{3}; perhaps it should now be learned or relearned by others.

\end{document}